\documentclass[
    ,final            
  ]
  {aipproc}

\layoutstyle{6x9}

\begin{document}

\title{Cosmological phase space of $R^n$ gravity}


\author{Alejandro Aviles Cervantes}{
  address={Instituto de Ciencias Nucleares,  Universidad Nacional
Aut\'onoma de M\'exico, A.P. 70-543,  04510 M\'exico D.F.,
M\'exico.}
}

\author{Jorge L. Cervantes-Cota}{
  address={Depto.  de F\'{\i}sica, Instituto Nacional de Investigaciones Nucleares,
A. P. 18-1027, M\'exico D.F. 11801, M\'exico.}
}

\begin{abstract}
We present some exact solutions and a phase space analysis of metric $f(R)$--gravity models of the type 
$R^{n}$.  We divide our discussion in $n\neq2$ and  $n=2$ models. The 
later model is  a good 
approximation, at late times to the $f(R) = \frac{2}{\pi}R \tan^{-1}(R/\beta^2)$ gravity model, being this an example of a  
non--singular case.   For   $n \neq 2$  models we have found power law 
solutions for the scale factor that are attractors 
and that comply with  WMAP 5-years data if  $n <-2.55\,$ or $\,1.67< n < 2$.  On the other hand, the quadratic model has 
the de Sitter solution as an  attractor, that also complies with   WMAP 5-years data.
 \end{abstract}

\maketitle


\section{Introduction}
The standard model of cosmology (SMC), which is based upon the Friedmann-Robertson-Walker (FRW) models in the 
context of  General Relativity (GR),  is successfully to explain the main dynamical and kinematical features of our 
Universe, such as the stretching of light curves  of type Ia supernovae  at high redshifts ($z \sim 0.3-1$) \cite{Ri98, Pe99, Fi04},  
CMBR anisotropy measurements \cite{Sp07,WMAP5y},  and structure formation features \cite{Pe02,Ef02}, among other tests.  To 
achieve all this it is necessary to  introduce two unknown universe's components: dark matter (DM) and dark energy (DE).   It is 
an unfortunate fact that we do not yet have any proved particle physics 
model that can  account for these dark 
components, though some good candidates exist. 

In the SMC one employs the Einstein-Hilbert  action
\begin{equation} \label{eins-hilb}
S_{EH} = \int \frac{1}{2 \kappa}R \sqrt{-g} d^4 x, 
\end{equation}
where $R$ is the Ricci scalar of the space-time, $g$ is the determinant 
of the metric and $\kappa = 8 \pi G$.  This gives the well known 
Einstein field equations

\begin{equation}
G_{\mu\nu} \equiv R_{\mu\nu} - \frac{1}{2} \, g_{\mu\nu} R  = \kappa 
(T_{\mu\nu} + T_{\mu\nu}^{(DE)}),
\end{equation}
where $T_{\mu\nu}$ includes all the known matter fields and,  if
necessary, the unknown DM.  Positive acceleration is possible in virtue of  the 
$T_{\mu\nu}^{(DE)}$ tensor, which  might be a cosmological constant
or a scalar field, generically called quintessence  \cite{Caldwell}, to which one imposes the condition 
on its pressure to be negative, $ p < - (1/3) \, \rho $.  The {\it technology} known for the inflationary dynamics can then 
be applied to the recent cosmological acceleration \cite{CoSaTs06}.

Alternatively,  in recent years it has appeared in the literature a new 
kind of 
models, generically called $f(R)$, which were initially aimed to account 
for the present acceleration of the 
universe \cite{Ca02, Carrol}, hence avoiding de need of DE.  It is 
possible that the above--mentioned observations indicate that GR 
needs to be modified at low energy scales that cosmologically correspond to recent times.  More recently 
these models have been used  to explain rotation curves \cite{Ca04,MaSa07} and cluster dynamics 
without  DM \cite{CaDeSao8}.  

With the course of the years many models and properties have been published in which different fundamental variables 
has been chosen to vary the lagrangian, i.e., apart of the metric, one can also choose the Palatini formalism in which 
the connections are field variables as well, a formalism called {\it Palatini} $f(R)$ {\it gravity}, in contrast to 
the {\it metric} $f(R)$ {\it gravity} that only uses the metric.   Both of these formalisms use matter fields independent 
of the connections, but one 
is not forced to do it so. Matter fields that possibly depend on the connections are 
called  {\it metric-affine} $f(R)$ {\it gravity}; for more on this topic 
see \cite{So07,SoFa08}.  The present paper 
assumes the standard   {\it metric} $f(R)$ {\it gravity} to analyze some 
exact solutions and to study the phase space.  First, we study $R^n$ 
gravity models with $n \neq 2$ and, secondly, we consider a 
non-singular $f(R)$ model that at late times  is well approximated by $R^2$.

\section{$f(R)$ field equations}

In order to geometrically modify GR, one can use some of the known 
gravity invariants, and among  all possible ones, the   
simplest modification is to replace the lagrangian  (\ref{eins-hilb}), 
$L = R$, with a generic function of the scalar 
curvature \cite{Eddington, Magnano1}, $f(R)$. Thus, the modified gravity action is  generically written as

\begin{equation} \label{AfR}
S = \int \frac{1}{2 \kappa} f(R) \sqrt{-g} \, d^4 x.
\end{equation}

By performing the variation of the action one obtains the field 
equations of $f(R)$ theories \cite{Buchdahl}:

\begin{equation} \label{EcCa}
f'(\!R) R_{\mu\nu} - \frac{1}{2} \, g_{\mu\nu} f(\!R) \, +
(g_{\mu\nu} \nabla^{\rho}
   \nabla_{\rho}-\nabla_{\mu}\nabla_{\nu})f'(\!R) = \kappa \,T_{\mu\nu}.
\end{equation}

This is a coupled nonlinear partial differential equations system
of fourth order in the metric. The trace of equation (\ref{EcCa}) is

\begin{equation} \label{EcCaT}
3 \nabla^{\sigma} \nabla_{\sigma} \,f'(R) + R f'(R) - 2f(R) =
\kappa T,
\end{equation}
while the trace equation of GR is $R = -  \kappa \, T$. This is a
substantial difference,  since equation (\ref{EcCaT}) is a dynamical law
of the scalar curvature, while its counterpart in GR is an
algebraic law. Thus, we have introduced a new dynamical degree of
freedom as we have changed the lagrangian from $R$ to
$f(R)$.

The flat cosmological  Friedmann equations for $f(R)$ theories 
are, using $R= 6(\dot{H} + 2 H^2) = 0$,

\begin{equation} \label{F1FR}
\frac{1}{2}f - 3 (\dot{H} + H^2)f' + 18(\ddot{H}H +
4\dot{H}H^2)f'' = \kappa \, \rho
\end{equation}
and

\begin{equation} \label{F2FR}
(\dot{H} + 3H^2) f' - \frac{1}{2} f -2 H \partial_0 f' -
\partial_0\partial_0 f' = \kappa \,p.
\end{equation}

In the following we analyze these equations  for some $f(R)$ models. 

\section{$R^n$ Models}
First, we want to analyze the simplest possible models in the
$f(R)$ theories context, which are power laws of the scalar
curvature,

\begin{equation} \label{LRn}
f(R) = b^{-2(n-1)} R^n,
\end{equation}
where $b$ is a constant with units of mass and $n$ is a nonzero real
number. By using the Friedmann equation (\ref{F1FR}) we obtain

\begin{eqnarray} \label{ecFRn}
\frac{b^{2(n-1)} \,6^n}{2}(\dot{H} +
2H^2)^{n-2}\!\!\!&\!&\!\!\!\!\!\!\! \{ n(n-1)\ddot{H}H +
(1-n)\dot{H}^2 + \nonumber\\ & &\!\!\!\!\!\!(4n^2 - 7n
+4)\dot{H}H^2 +2(2-n) H^4\} = \kappa\rho.
\end{eqnarray}

First we note that the case $R= 6(\dot{H} + 2 H^2) = 0$ is
meaningless for $n<2 \,(n \neq 1)$, in contrast to GR where $R=0$
in vacuum or for matter fields with $T=0$. In vacuum, with
$\dot{H} + 2 H^2 \neq 0$, equation (\ref{ecFRn}) reduces to

\begin{equation} \label{ecFRnV}
n(n-1)\ddot{H}H + (1-n)\dot{H}^2 + (4n^2 - 7n +4)\dot{H}H^2
+2(2-n) H^4 = 0.
\end{equation}

We seek for power law solutions $a(t)\propto t^{\alpha}$, $H= \alpha/t$.
Using equation (\ref{ecFRnV}) we obtain the equation $ 2(2-n) \alpha^2 -
(4n^2 - 7n +4)\alpha + 2n^2 -3n+1=0$.  One of its roots
 is $\alpha = 1/2$ which is a solution for any $n$, but it complies with 
 $\dot{H} + 2 H^2 = 0$  and for $n<2 \, (n \neq 1)$ it is not acceptable 
in this description.  The 
 other root is given by

\begin{equation} \label{solpotRnV}
\alpha = \frac{2n^2 - 3n +1}{2-n}.
\end{equation}

For an accelerated universe, $\alpha>1$, this equation implies 
$1\textrm{.}36 <  n < 2\,\,$ or $\,\, n < - 0\textrm{.}36$.  The 
solution (\ref{solpotRnV}) is GR equivalent 
with a dark energy perfect fluid component with an effective equation of state parameter given by

\begin{equation}
w_{eff} = -1 + \frac{2(n-2)}{3(2n-1)(1-n)} .
\end{equation}

The 5-year WMAP data constrains $w_{eff}$ to
$-0.11 < 1+w_{eff} < 0.14\,$ \cite{WMAP5y}. This implies that in a
power law model, with $n\neq2$, $n$ is constrained to $n <
-2.55\,$ or $\,1.67< n < 2$. Note as well that $w_{eff} > -1$
can only be achieved for a power law expanding universe. 


By making a Legendre map of the lagrangian density from the nonlinear frame \cite{Magnano} -also 
called matter frame \cite{Carrol}- to the Einstein frame, Carrol et al.
 showed that a term $1/R^{m}$ in the lagrangian can
drive an accelerated expansion $a(t) \propto
t^{(2m+1)(m+1)/(m+2)}$ \cite{Carrol}, which is in perfect agreement with
equation (\ref{solpotRnV}), where $n=-m$.

\begin{figure} \label{fig1}
    \includegraphics[height=0.4\textheight]{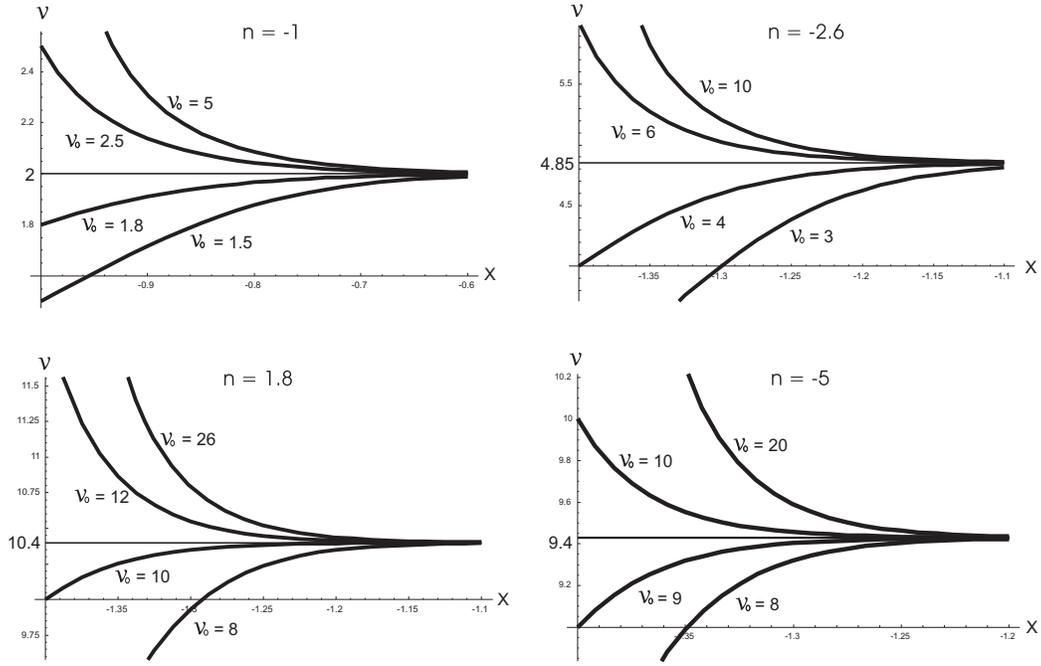}
\caption{Plots $v$ vs $x$.  Here we show results with $v_0 > 1/2$, which asymptotically tend to power law solutions that are given by $v(x) = $const.  \label{fRn2} }
\end{figure}

These power law solutions are attractors, as it can be shown by defining the variables 
$x=-H(t)$, $y=\dot{H}(t)$ and $v(x)=-x^2/y$ \cite{St80,Carrol2}.  Equation (\ref{ecFRnV}) then
reads

\begin{equation} \label{ecFRn3}
n(n-1)\frac{x^5}{v^3} \frac{dv}{dx} - (2n^2 - 3n + 1)
\frac{x^4}{v^2} + (4n^2 -7n +4)\frac{x^4}{v} + 2(n-2)x^4 = 0
\end{equation}
and an asymptotic behavior to a power law is identified with
$v(x)\rightarrow \alpha = \textrm{constant}$, as $H \rightarrow 0$. By performing
numerical integrations for different initial conditions $v_0$ we
obtain the plots shown in Fig. \ref{fig1} for  $n = -1$ and for various values of $n$ that are in the 
interval allowed by the WMAP data mentioned above.   Accelerated solutions are found for $v_0 > 1/2$.
Note that $v_0 = 1/2$ corresponds to $R=0$. It can be shown that if
$v_0 < 1/2$, the solutions tend  to $v=0$, \textit{i.e.}
$\dot{a}(t) \rightarrow 0$.

\section{Other Models}
There exist a plenty of $f(R)$ proposals \cite{SoFa08}, but have to 
be analyzed in order to verify they pass standard 
gravitational tests, such as stability criteria or to have a reliable Newtonian limit.  For instance, theories with $f''(R)<0$ and $|f''(R)|\ll 1$, 
such as $1/R^m$ or $\ln R$, are ruled out due to instabilities that appear when trying to model gravity fields in stars \cite{Seifert}.  

The initial $f(R)$ models, and some others, add to the lagrangian $L = 
R$  singular $R-$functions  such as $-1/R^{m}$ (with $m > 0$) 
having a difficult interpretation to obtain a Newtonian limit, since 
as $R \rightarrow 0$ the theory diverges.  We think  a less restrictive type of theories is to consider $m < 0$ that are 
non--singular in this sense.  

Motivated by the these facts we propose the following model:

\begin{equation} \label{modarctan}
f(R) = \frac{2}{\pi}R \tan^{-1}(R/\beta^2),
\end{equation}
in which $\beta$ is a constant with units of mass.  Indeed, this model passes the 
stability criteria and has reliable Newtonian limit; these matters 
will be treated in a separate work.  For $R \gg \beta^2$ the model behaves basically as $f(R) \rightarrow R -
\frac{2}{\pi} \beta^2$, which is GR with a cosmological constant.
We define $t_{\beta}$ as the time when $R\sim
\beta^2$ and we have a transition period. To adjust to
observations we demand $\beta \sim H_0 \sim 10^{-33} eV$.

At late times, $R<\beta^2$,  the model is well approximated by
$f(R) \approx 2R^2/ \pi\beta^2$. Using the Friedmann equation
(\ref{F1FR}) in vacuum we obtain

\begin{equation} \label{arctanedv}
2\ddot{H}H + 6\dot{H}H^2 - \dot{H}^2 = 0.
\end{equation}

\begin{figure}[ht]  \label{fig2}
    \includegraphics[height=0.4\textheight]{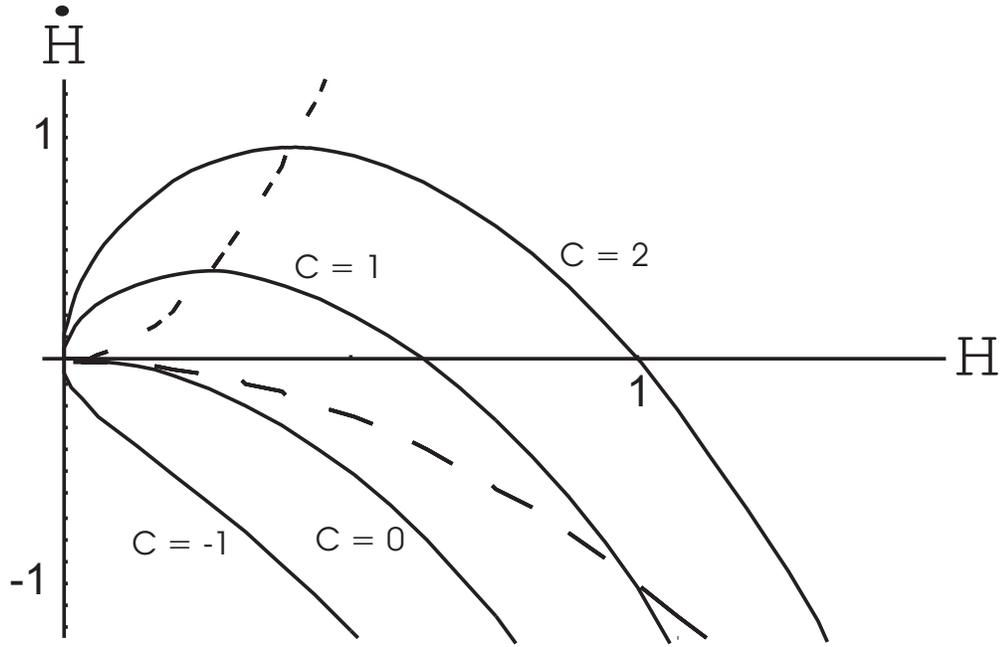}
\caption {Phase space showing accelerated solutions for curves that are  above the 
long--dashed curve.  \label{fEF}}
\end{figure}

By integrating once, it yields,

\begin{equation} \label{arctanedv2}
\dot{H} +2H^2 - C H^{1/2} = 0,
\end{equation}
with $C$ an integration constant. If $C=0$ we obtain $a\propto
t^{1/2}$ and $a =$ const, which are not interesting for our aim. In  
Fig. \ref{fig2} the phase space 
is shown for  several values of $C$. The long--dashed curve is the
equation $\dot{H} = -H^2$, and the acceleration condition is
accomplished in the region where the solutions are above it. The double--dashed 
curve, $\dot{H}= 6 H^2$, is the locus of the maximae 
which occur for all the solutions with $C>0$.   Accelerated
solutions occur only for $C>0$.   The initial condition 
problem is to specify $\dot{H}_0$ and $H_0$ and then, by using (\ref{arctanedv2}), one 
determines $C = (\dot{H}_0 + 2 H^2_0)/H^{1/2}_0$ to solve the dynamics.

\begin{figure}[ht]  \label{fig3}
\includegraphics[height=0.4\textheight]{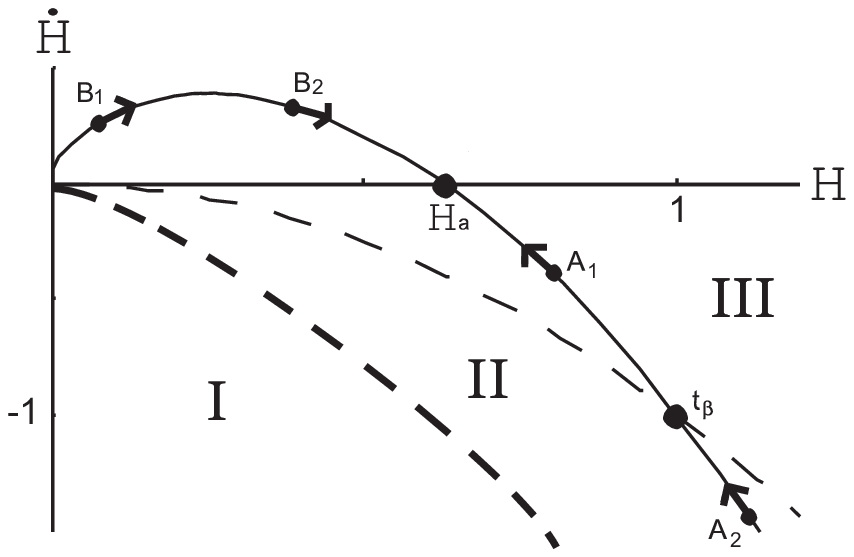}
\caption{Phase space for $C=1$.  Notice that $H_a$ is an attractor to 
all 
solutions with $C>0$.  \label{fHHp}}
\end{figure}

In Fig. \ref{fig3} we show the solution with $C=1$.  In general for $C>0$,  a phase point 
\textbf{A} with $H_0>0$ and $\dot{H}_0<0$ complies with $\ddot{H} > 0$, as it follows from
equation (\ref{arctanedv}), and the solution
evolves to the phase point $H_a$. For a point \textbf{B} with
$H_0>0$ and $\dot{H}_0>0$, $\ddot{H} > 0$ if  $\dot{H}> 6H^2$ and
$\ddot{H}< 0$ if  $\dot{H}< 6H^2$. That is, for all $C>0$, the
solutions evolve to $H \rightarrow H_a = C^{2/3} / 4^{1/3}$, that corresponds to 
an exponential de Sitter solution.

We can write equation (\ref{arctanedv2}) as $R= 6 C H^{1/2}$, and we 
point out 
that if $R > 0$ the solution will have an accelerated final stage.
Thus we divide the phase space in three regions:

\textbf{I}: $R<0$; The solutions are never accelerated and tend to
$a(t) \rightarrow a_0 = \textrm{cte.}$

\textbf{II}: The region between $R=0$ ($\dot{H} = -2 H^2$) and
$\dot{H} = -H^2$. Here solutions are not accelerated but they
reach $t_{\beta}$ in a finite time and enter region III.

\textbf{III}: The solutions are asymptotically de Sitter, $a(t)
\rightarrow a_0 e^{H_a t}$.

In order to adjust to observations we have $H_a \sim H(t_{\beta}) \sim 
H_0 \sim \beta
\sim 10^{-33} eV$, and from $H(t_{\beta}) = C^{2/3}$ it follows that $C
\sim \beta^{3/2} \sim 10^{-51} eV^{3/2}$.

Summarizing, the \textit{Arctangent} model describes a universe which
initially evolves such as the one described by GR with a cosmological
constant into a stage dominated by a quadratic term of the scalar
curvature.  In this stage the solutions tend to a de Sitter space-time.




\section{Conclusions}
In the phase space analysis presented in this contribution we have analyzed $R^{n}$ and the $Arctangent$ model. For the
former model with $n \neq 2$ we have found power law solutions that 
resulted to be attractors and that comply with the constraints 
imposed by WMAP \cite{WMAP5y} if  $n < -2.55\,$ or  $\,1.67< n < 2$, yielding $w_{eff} > -1$.     The $Arctangent$ model, that was 
motivated as an example of a non-singular $f(R)$-gravity, is approximated at late times as a quadratic function of $R$.  The resulting 
equations can be  once integrated, leading to the simple equation 
(\ref{arctanedv2}) with a constant ($C$) that depends on the initial 
conditions ($H_0, \, \dot{H}_0$).  We have found three phase space regions having the following properties: In region I, that corresponds  
to negative values of $C$ and hence to $R<0$, the solutions are non-accelerating.  In region II, delimited by two parabolic functions, solutions
are not accelerated but they tend to region III in a finite time of the order of $t_{\beta}$, where solutions are accelerated and tend to a de Sitter
solution. Therefore, regions II and III of the phase space end with $w_{eff} = -1$, that is also in accordance with the WMAP results.

\begin{theacknowledgments}
This work has been supported by CONACYT, grant number 84133 and PhD scholarship 215819.
\end{theacknowledgments}


\end{document}